\documentclass[a4paper]{jpconf}
\usepackage{graphicx}
\usepackage{textcomp}
\usepackage{hyperref}
\usepackage{amssymb,amsmath}
\begin{document}
\title{Dissipative Dynamics of a Single Polymer in Solution: A Lowe-Andersen Approach}

\author{Suman Majumder, Henrik Christiansen and Wolfhard Janke}

\address{Institut f\"ur Theoretische Physik, Universit\"at Leipzig, 
Postfach 100\,920, 04009 Leipzig, Germany}

\ead{suman.majumder@itp.uni-leipzig.de \\~~~~~~ henrik.christiansen@itp.uni-leipzig.de \\ 
~~~~~~~wolfhard.janke@itp.uni-leipzig.de}

\begin{abstract}
We study the equilibrium dynamics of a single polymer chain under good solvent condition. 
Special emphasis is laid on varying the drag force experienced by the chain 
while it moves. To this end we model the solvent in a mesoscopic manner by employing 
the Lowe-Andersen approach of dissipative particle dynamics which is known to 
reproduce hydrodynamic effects. Our approach captures the correct 
static behavior in equilibrium. Regarding the dynamics, we investigate the 
scaling of the self-diffusion coefficient $D$ with respect to the length of 
the polymer $N$, yielding results that are compatible with the Zimm scaling 
$D \sim N^{-3/5}$. 
\end{abstract}
\section{Introduction}
Dynamics of a polymer chain in a dilute solution, although being extensively 
studied, is still a topic of utmost importance \cite{deGennes_book,Doi_book}. In 
particular, this topic serves as a benchmark for establishing a coarse-grained or mesoscopic 
approach to understand more realistic problems on larger time and length scales. 
The dynamics of a single chain, generally, is characterized by the self-diffusion 
coefficient $D$ which scales with chain length $N$ as
\begin{equation}\label{Dscaling}
 D \sim N^{-x}.
\end{equation}
In the free-draining limit where hydrodynamic effects are absent or screened, one has $x=1$, 
whereas in the non-draining limit where hydrodynamic interactions are important, one expects 
$x=3/5$. The former is referred to as Rouse scaling \cite{Rouse} and the latter 
as Zimm scaling \cite{Zimm}. 
\par
Over the years these results have been verified both experimentally \cite{Smith} as 
well in computer simulations \cite{Dunweg}. Before the advent of present-day sophisticated 
experimental setups, when it was difficult to keep track of a single polymer movement, 
numerical simulations were considered to be the only way to verify the available theoretical 
understanding. Monte Carlo simulations cannot capture the real 
dynamics due to the absence of hydrodynamic effects. Even when doing Molecular Dynamics (MD) 
simulations, one has to be careful about the choice of the thermostat. The effect of 
hydrodynamics is achieved via the preservation of local linear and angular momentum during the entire 
simulation. 
\par
In the context of polymer dynamics another major issue is the consideration of explicit solvents. 
The introduction of dissipative particle dynamics (DPD) has eased this task 
\cite{Koelman}. There one has the luxury of considering the solvent explicitly via a 
mesoscopic approach which allows to access larger time and length scales \cite{Espanol,Groot}. 
In addition, DPD also allows to tune the Schmidt number, i.e., the ratio of the 
kinematic viscosity and the self-diffusion coefficient. This makes consideration of solvents 
having viscosities comparable to real fluids quite plausible. However, from a technical point of view, 
one has to be cautious when integrating the equations of motion using a Verlet-type 
algorithm in the DPD formalism as it often disturbs the detailed balance and thereby does not produce 
the correct equilibrium properties unless sufficiently small time steps are used \cite{Yeomans,Liew}. 
There have been few attempts to use more advanced integration schemes that can overcome this difficulty. 
In this context, Lowe instead of aiming at improving the integration scheme, modified the DPD 
approach in the spirit of the Andersen thermostat, of course, with the effect of hydrodynamic 
interactions being intact \cite{Lowe}. With Lowe's approach, which also goes by the name Lowe-Andersen (LA) 
thermostat, one is not only allowed to use relatively larger time steps \cite{Koopman}
but also has the option to tune the dissipation of velocities of particles, thereby gaining access 
to fluids with Schmidt numbers as high as observed in real fluids \cite{Lowe}. Motivated by these advantages, 
here, we construct a model polymer with explicit solvent particles and perform MD simulations  
for a wide range of effective solvent viscosities, i.e., frictional-drag experienced by the particles. Our results 
show that the dynamics in good solvent produces the theoretically expected Zimm scaling valid in the 
presence of hydrodynamics.
\par
The paper is organized as follows. In the next section we present the details of our model 
and the method of simulation. Following that, we present results concerning static and dynamic 
properties of our model. In the final section we present a summary of our results and an outlook to future work.
\section{Model and Method} 
We consider a bead-spring model of a flexible homopolymer in three spatial dimensions. The 
bonds between successive monomers are maintained via the standard 
finitely extensible non-linear elastic (FENE) potential 
\begin{eqnarray}\label{FENE}
E_{\rm{FENE}}(r_{ii+1})=-\frac{K}{2} R^2 \ln \left[ 1-\left( \frac{r_{ii+1}-r_0}{R} \right)^2 \right],
\end{eqnarray}
with $K=40$, $r_0=0.7$ and $R=0.3$. The monomers and the solvent molecules both are considered to be spherical 
beads of mass $m=1$ and diameter $\sigma$. All nonbonded interactions, i.e., solvent-solvent, solvent-monomer, 
and monomer-monomer interactions are modeled by 
\begin{equation}
 E_{\rm {nb}}(r_{ij})=E_{\rm {LJ}}\left[{\rm{\min}}(r_{ij},r_c)\right]-E_{\rm {LJ}}(r_c),
\end{equation}
where $E_{\rm {LJ}}(r)$ is the standard Lennard-Jones (LJ) potential given as 
\begin{equation}
 E_{\rm {LJ}}(r)=4\epsilon \left [ \left(\frac{\sigma}{r} \right)^{12} - \left( \frac{\sigma}{r} \right)^{6} \right]
\end{equation}
with $\sigma =r_0/2^{1/6}$ as the diameter of the beads, $\epsilon(=1)$ as 
the interaction strength and $r_c$ $=2^{1/6}\sigma$ as cut-off radius that 
ensures a purely repulsive interaction. 
\par
As already mentioned, we simulate our system via MD simulations at constant temperature using the LA 
thermostat. In this approach, one updates the position $\vec{r}_i$ and velocity $\vec{v}_i$ 
of the $i$-th bead using Newton's equations as follows, 
\begin{equation}\label{newton}
 \frac{d\vec{r}_i}{dt}=\vec{v}_i, \frac{d\vec{v}_i}{dt}=\vec{f}_i,
\end{equation}
where $\vec{f}_i$ is the conservative force (originating from the bonded and nonbonded interactions) acting 
on the particle. This part of the simulation is the usual microcanonical MD. For controlling the temperature with 
the LA thermostat, one considers a pair of particles within a certain 
distance $R_T$. Then, with a probability $ \Delta t \Gamma$, a bath collision is executed following 
which the pair gets a new relative velocity from the Maxwellian distribution. Here, $\Delta t$ 
is the width of the time step chosen for the updates in Eq.\ \eqref{newton} and $\Gamma$ determines 
the collision frequency. The exchange of relative velocities with the bath is only done on 
its component parallel to the line joining the centers of the pair of particles, thus conserving the angular 
momentum. Additionally, the new velocities are distributed to the chosen 
pair in such a way that the linear momentum is also conserved. In summary, the work flow for our LA approach 
has the following form:
\begin{enumerate}\label{collision}
\item Update $\vec{v}_i$ \textleftarrow $\vec{v}_i +\frac{1}{2m}\vec{f_i} \Delta t$ and $\vec{r}_i$ \textleftarrow $\vec{r}_i +\vec{v}_i \Delta t$.
\item Calculate $\vec{f_i}$ and then $\vec{v}_i$ \textleftarrow $\vec{v}_i +\frac{1}{2m}\vec{f_i} \Delta t$.
\item Choose all pairs of particles with $r_{ij} < R_T$ and with probability $\Delta t\Gamma$ do the following:
\begin{enumerate}
\item Draw ${\vec{v}\prime}_{ij} \cdot \vec{n}_{ij}$ from the distribution $\xi_{ij} \sqrt{2k_BT/m}$ where ${\vec{v}\prime}_{ij}$ is the new relative velocity of particles $i$ and $j$, 
$\vec{n}_{ij}$ is a unit vector, $\xi_{ij}$ is a Gaussian white noise, and $T$ is the desired temperature.
\item Calculate the change $2\vec{\Delta}_{ij}=\vec{n}_{ij}({\vec{v}\prime}_{ij} - \vec{v}_{ij}) \cdot \vec{n}_{ij} $.
\item Distribute the change as $\vec{v}_i$ \textleftarrow $\vec{v}_i +\vec{\Delta}_{ij}$ and $\vec{v}_j$ \textleftarrow $\vec{v}_j -\vec{\Delta}_{ij}$.

\end{enumerate}
\item Calculate physical quantities and go to (i). 
\end{enumerate}
It is to be noted that the LA approach is an alternative approach to DPD, however, with the liberty to use large $\Delta t$. The other 
advantage of the method is that by varying $R_T$ and $\Gamma$ one can tune the bath collision frequency, i.e., effectively 
controlling the frictional drag or in other word the solvent viscosity. In this work, we restrict ourselves 
to the case where $R_T=r_c$ and vary $\Gamma$ within the range $[0.1,250]$ with the goal to cover solvents with diverse 
viscosity. We do our simulations using LAMMPS \cite{lammps} which we modified to implement the LA thermostat.
\par
We first generate a random walk of length $N$ on a simple-cubic lattice and then put this walk or chain in a 
box of size $L=1.25N^{3/5}$. Subsequently, we insert solvent particles keeping the density fixed to $\rho=0.7$ and make sure that the solvent particles do not overlap 
with the monomers of the polymer chain. Then we run our MD simulation with LA thermostat at temperature $T=1.0$ for $10^7$ MD steps with $\Delta t=0.005$ 
and allow the system to equilibrate. In our simulations, the unit of temperature is $\epsilon/k_B$ (where we chose $k_B=1$) and the unit of time 
is the standard LJ time unit $\tau=(m\sigma^2/\epsilon)^{1/2}$. Once the system is equilibrated, we let it run for another period 
of $10^4 \tau$ and simultaneously start measuring various physical quantities that will be presented subsequently. We 
have used polymers of chain length $N \in [16,512]$. All results presented are averaged over $100$ different independent 
runs for $N<512$ and $50$ runs for $N=512$. 
\begin{figure}[t!]
\centering
\includegraphics*[width=0.4\textwidth]{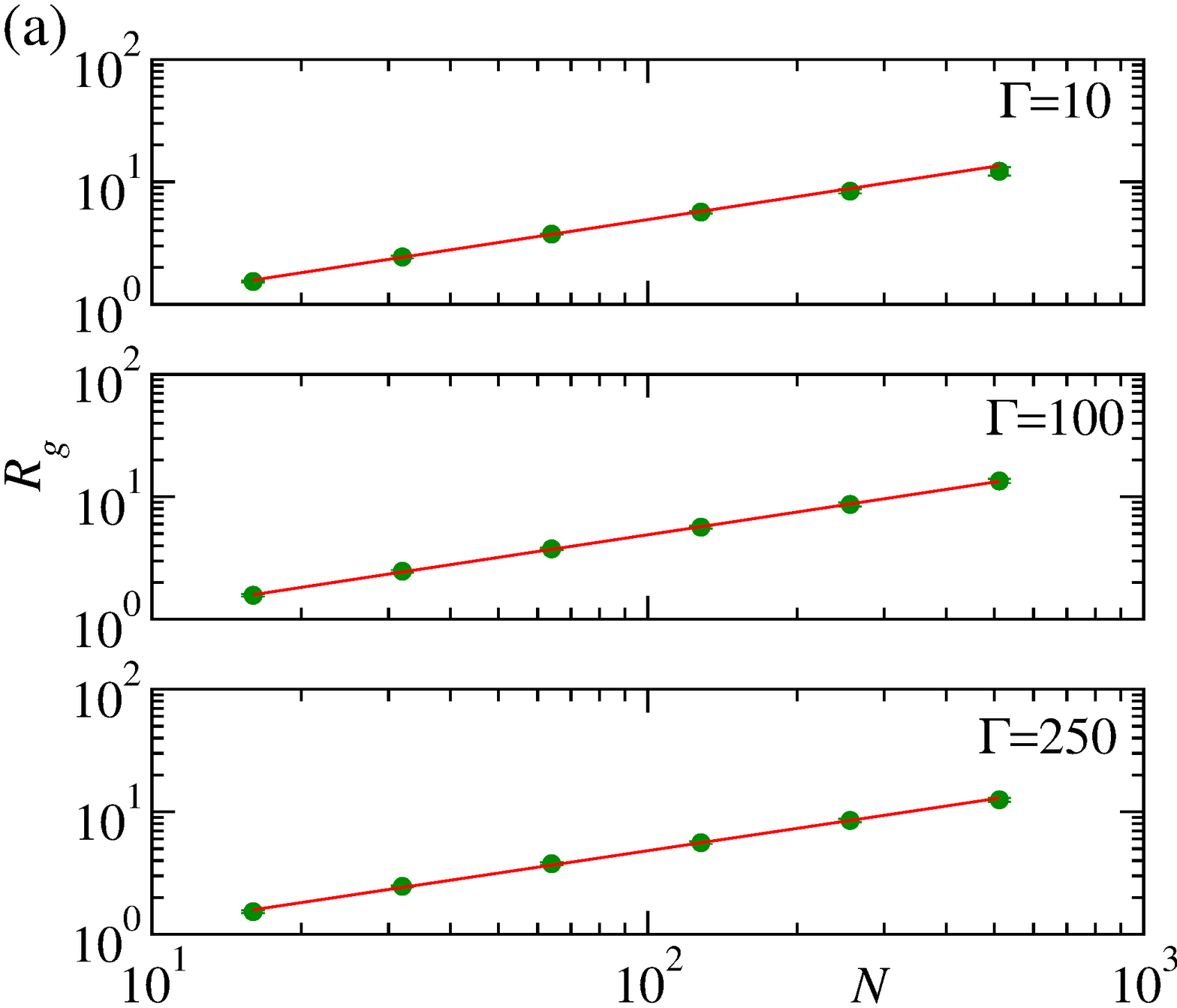}
~~\includegraphics*[width=0.5\textwidth]{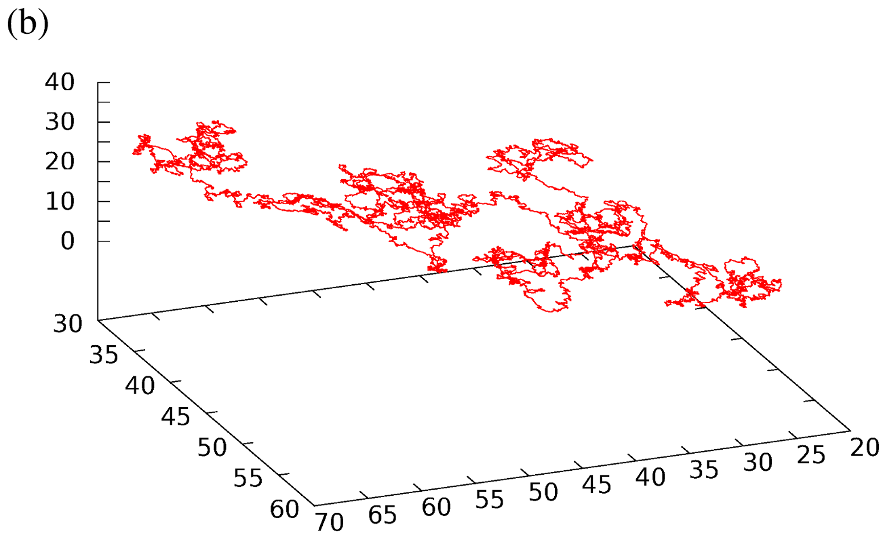}
\caption{\label{fig1}(a) Radius of gyration $R_g$ as function of the chain length $N$ in good solvents for three different $\Gamma$. The 
solid line in each case is a fit using the form $R_g=R_0 N^{3/5}$, the expected scaling behavior. (b) The trajectory of the center of mass 
of a polymer of length $N=512$, over a time period of $200 \tau$ in equilibrium at $T=1.0$.}
\end{figure}
\section{Results}\label{Results}
As a first step to benchmark our proposed framework we calculate the radius of gyration as 
\begin{equation}
 R_{g}=\sqrt{\sum\limits_{i,j}\frac{(\vec{r}_i-\vec{r}_j)^2}{2N^2}},
\end{equation}
which is a measure for the size of the polymer. Under good solvent 
condition, this scales with the chain length $N$ as $R_g \sim N^{\nu}$ where the critical exponent $\nu \approx 3/5$. In 
Fig.\ \ref{fig1}(a) we show the plots of $R_g$ as function of $N$ for three different $\Gamma$ values of the solvent. 
A fitting of the data using the form $R_g=R_0N^{\nu}$ provides $\nu \in [0.58,0.61]$. Fixing $\nu=3/5$ in the fitting, also 
works quite well as shown by the continuous lines in Fig.\ \ref{fig1}(a). This implies that our framework reproduces the correct 
equilibrium static behavior of a polymer in good solvent, irrespective of the value of $\Gamma$.
\par
\begin{figure}[t!]
\centering
\includegraphics*[width=0.44\textwidth]{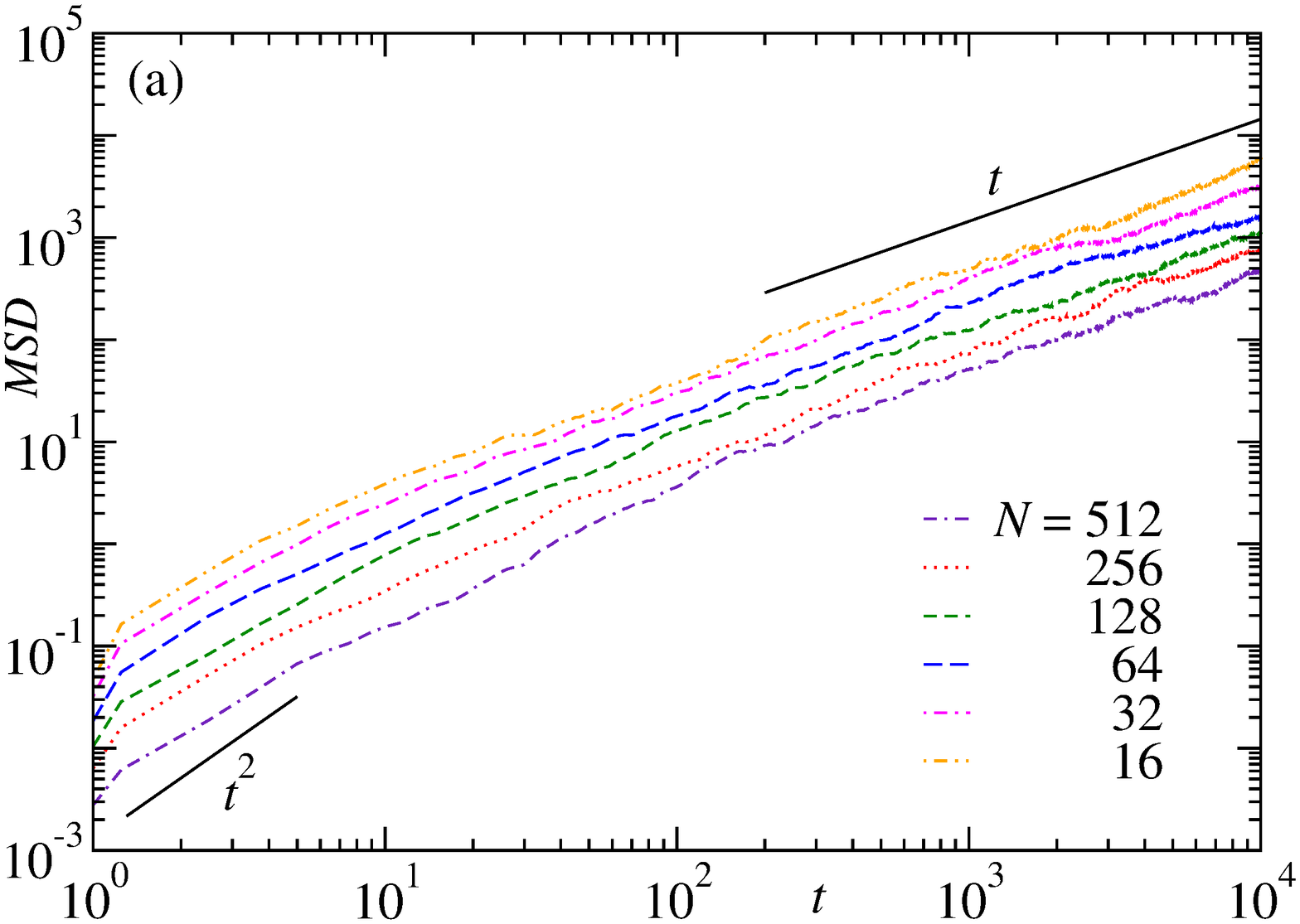}
~~\includegraphics*[width=0.44\textwidth]{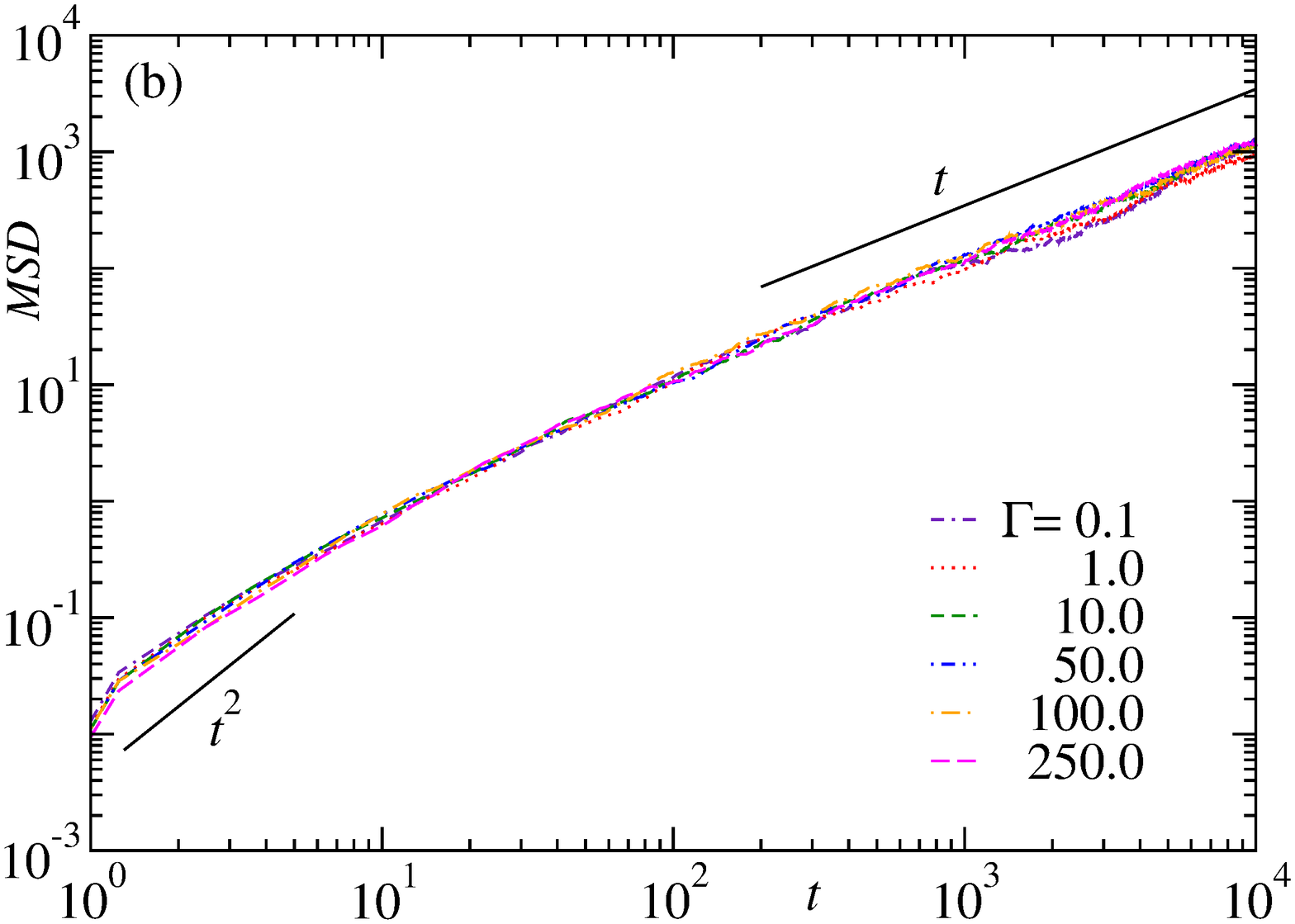}
\caption{\label{fig2} (a) Time dependence of mean squared displacement $MSD$ of the center of mass of a polymer for 
different chain lengths $N$ in a solvent having $\Gamma=10$ at $T=1.0$. The solid line at small $t$ corresponds to ballistic motion $\sim t^2$ and the 
one at large $t$ corresponds to Brownian motion $\sim t$. (b) Same as (a) but for a polymer of fixed length $N=128$ in solvents 
having different $\Gamma$. The solid lines there have the same meaning as in (a).}
\end{figure}

Next we move to the dynamic properties. In Fig.\ \ref{fig1}(b), we show the equilibrium trajectory of the center of 
mass (cm) of a polymer of length $N=512$ over a time interval of $200 \tau$. Tracking the motion of the cm of a polymer is 
crucial when one wants to measure its diffusion in the solvent. The trajectory seems to be stochastic in nature and hence is 
indicative of a Brownian motion. To check the nature of the motion we calculate the mean squared displacement of the 
cm of a polymer given as 
\begin{eqnarray}\label{MSD}
MSD= \langle [\vec{R}_{\rm{cm}}(t)-\vec{R}_{\rm{cm}}(t_0)]^2 \rangle,
\end{eqnarray}
where $\vec{R}_{\rm{cm}}(t)$ is the position vector of the cm of the polymer at time $t$, and $t_0$ is the time when the measurement starts. From Einstein's equation 
it is known that for Brownian motion in the long-time limit \cite{deGennes_book,Doi_book}
\begin{equation}\label{Einstein}
 MSD=A+6Dt,
\end{equation} 
where $D$ is the self-diffusion coefficient of the polymer and $A$ is a constant. In Fig.\ \ref{fig2}(a) 
we show the plots of $MSD$ as function of time $t$ for different chain lengths $N$ in a solvent with $\Gamma=10$. In the large $t$ limit, 
for all $N$ the data is consistent with $\sim t$ behavior, whereas at early times for a brief period it is expectedly ballistic 
in nature, i.e., $\sim t^2$. A similar behavior is observed in Fig.\ \ref{fig2}(b) where we show plots of $MSD$ for polymers having fixed chain length 
$N=128$ in solvents having different $\Gamma$.  
\par
To have a more comprehensive understanding of the effect of variation of the solvent, i.e., $\Gamma$, we aim to 
calculate the self-diffusion coefficient $D$ of the polymer. In this regard, one can calculate the velocity (of the cm of the polymer) 
autocorrelation function $C(t)= \langle \vec{v}(t_0) \cdot \vec{v}(t) \rangle$ 
which is related to $D$ via the Green-Kubo relation as 
\begin{equation}
 D=\frac{1}{d}\int_{t_0}^{t} C(t) dt,
 \end{equation}
where $d$ is the spatial dimension. However, here, we calculate $D$ using the Einstein relation \eqref{Einstein} as follows. We pick 
two times $t_1$ and $t_2$ in the long-time limit. From Fig.\ \ref{fig2}, one can easily observe that for $t > 100 \tau$ for all $N$ and $\Gamma$, 
the mean squared displacement $MSD$ is consistent with the linear behavior. Thus, we chose $t_1$ and $t_2$ to be such that $t_2>t_1$ and $t_1 \ge 100 \tau$. Then 
using \eqref{Einstein} one can write down 
\begin{equation}\label{Dcoff}
 D= \frac{1}{6}\left[\frac{MSD(t_2)-MSD(t_1)}{t_2-t_1} \right]. 
 \end{equation}
Equation\ \eqref{Dcoff} provides a set of values of $D$ for different choices of the pair ($t_1,t_2$). This allows one to have appropriate 
error bars, independent of the usual fitting exercise using the form \eqref{Einstein}. 
\par
\begin{figure}[t!]
\centering
\includegraphics*[width=0.43\textwidth]{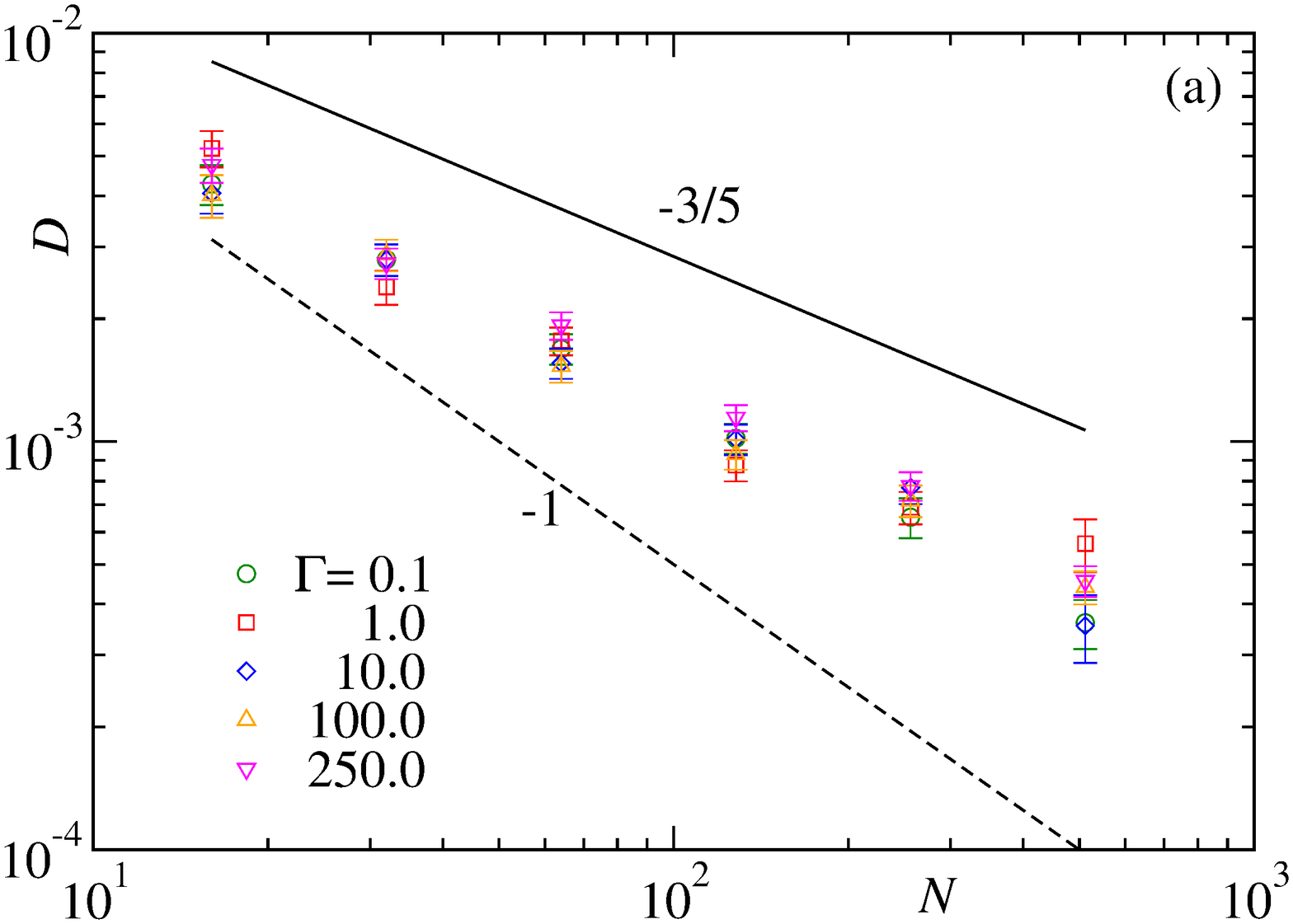}
~~~\includegraphics*[width=0.44\textwidth]{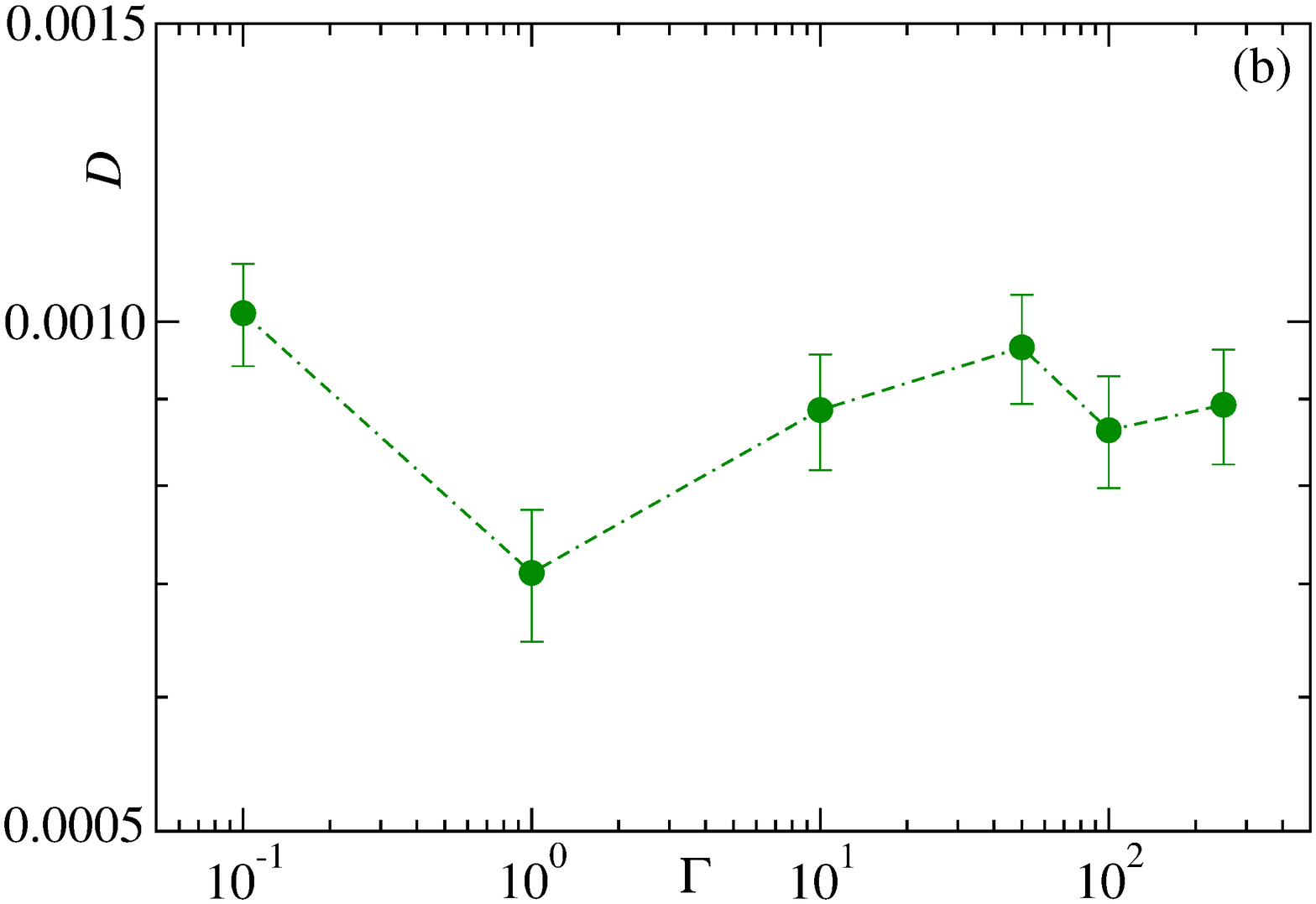}
\caption{\label{fig3} (a) Plots showing the dependence of the self-diffusion coefficient $D$ on the chain length $N$
of a polymer for different solvents with $\Gamma$ as indicated. The solid line and dashed lines represent respectively the Zimm and Rouse scaling. 
(b) Variation of $D$ as function of $\Gamma$ for a polymer of length $N=128$ at $T=1.0$.}
\end{figure}

Figure\ \ref{fig3}(a) demonstrates the scaling of the self-diffusion coefficient $D$ with chain length
$N$ for different solvents as indicated by the $\Gamma$ values therein. The dashed line there corresponds 
to the Rouse scaling with the exponent $x=1$ in the form \eqref{Dscaling} whereas the solid line represents the Zimm scaling 
with $x=3/5$.  Consistency of our data with the solid line indicates the validity of the Zimm scaling, expected in the 
presence of hydrodynamic effects. Fits using the form $D=D_0N^{-x}$ yield $x$ within $[0.6,0.68]$, slightly higher than 
$x=3/5$ expected for the Zimm scaling. However, a fit using the same form by fixing $x=3/5$ also yields reasonably acceptable 
$\chi_r^2 (< 2.5)$ values, where $\chi_r^2$ is the goodness of fit parameter $\chi^2$ divided by the degrees of freedom.
\par
Finally, in Fig.\ \ref{fig3}(b) we show the dependence of the self-diffusion coefficient $D$ on the collision frequency $\Gamma$ 
that controls the effective viscosity of the solvent. The data 
show no signature of strong dependence, rather in a broader sense $D$ seems to be pretty constant for different $\Gamma$. This behavior 
is similar to the conclusion drawn in Ref.\ \cite{Koopman} where, for an ideal gas, they did not observe any strong dependence of $D$ 
on $\Gamma$. Nevertheless, the data presented here are 
for a relatively high temperature $T=1.0$, hence, the influence of viscosity is not severely pronounced. At low temperatures, 
the diffusion might get suppressed due to high effective viscosity \cite{Suman1}.
\section{Conclusion}
Motivated by dissipative particle dynamics, here, we have constructed an explicit solvent model for 
a polymer. Instead of using the standard approach of doing it, we rely on the Lowe-Andersen 
approach of the bath collision. This allows us to control the drag force 
applied by the solvent on the polymer, i.e., the effective viscosity. Via the scaling of the radius of gyration $R_g$ with the chain length $N$ as 
$R_g \sim N^{3/5}$ we confirm that our approach yields the known static critical exponent. The method 
conserves both the linear and angular momenta locally, thereby preserving the hydrodynamics. The scaling of the self-diffusion coefficient $D$ 
with chain length $N$ indicates a much faster dynamics than the Rouse dynamics, and in fact is pretty consistent 
with Zimm scaling $D \sim N^{-3/5}$ valid in the presence of hydrodynamic effects. The successful application of this method 
opens up opportunities to explore other aspects of polymer dynamics including nonequilbrium scenarios, e.g., during 
collapse of a polymer \cite{deGennes1,Suman_EPL,Suman_SM,Henrik}. There, the effect of varying the collision frequency $\Gamma$ 
has a stronger influence on the structure formation, and in principle such phenomena could well be tuned by the 
degree of dissipation \cite{Suman2}.
\ack
The work was funded by the Deutsche Forschungsgemeinschaft (DFG) under Grant Nos.\ JA 483/33-1 and SFB/TRR 102 (project B04), 
and further supported by the Deutsch-Franz\"osische Hochschule (DFH-UFA) through the Doctoral College ``${\mathbb L}^4$'' under Grant No.\ CDFA-02-07, 
the EU Marie Curie IRSES network DIONICOS under Contract No.\ PIRSES-GA-2013-612707, and the Leipzig Graduate School of Natural Sciences ``BuildMoNa''.

\end{document}